\documentstyle[12pt]{article}
\def\abstract#1{{\centerline{\bg Abstract}} \vskip 3mm \par #1}
\def\cy{Calabi-Yau}
\def\cym{Calabi-Yau manifold}
\def\lg{Landau-Ginzburg}
\def\lgo{Landau-Ginzburg orbifold}

\def\tg{\tilde{\theta_{i}}^{g}}

\def\thd{\tilde{\theta_{i}}^{h^{\prime}}}

\def\inbar{\vrule height1.5ex width.4pt depth0pt} 
\def\ZZ{\relax{\sf Z\kern-.4em \sf Z}}  \def\IR{\relax{\rm I\kern-.18em R}}
\def\IN{\relax{\rm I\kern-.18em N}} \def\IP{\relax{\rm I\kern-.18em P}}
\def\IQ{\relax\,\hbox{$\inbar\kern-.3em{\rm Q}$}}
\def\la{\lambda}
\def\IC{\hbox{\,$\inbar\kern-.3em{\rm C}$}}

\def\ket#1{\left| #1\right\rangle}

\def\({\lbrack}
\def\){\rbrack}

\def\ketc#1{{\left| #1\right\rangle}_{\rm (c,c)}}
\def\keta#1{{\left| #1\right\rangle}_{\rm (a,c)}}

\begin{document}
\baselineskip=6mm
\begin{flushright}
{hep-th/9507075} \\
{KOBE-TH-95-02} \\
{July 1995}
\end{flushright}
\vskip 1.5cm
\centerline{\LARGE {\bf Dual Polyhedra, Mirror Symmetry}}
\vskip 1cm
\centerline{\LARGE {\bf and}}
\vskip 1cm
\centerline{\LARGE {\bf Landau-Ginzburg Orbifolds}}
\vskip 1.5cm
\centerline{\large Hitoshi \ Sato}
\vskip 1cm
\centerline{\it Graduate School of Science and Technology, Kobe University}
\centerline{\it Rokkodai, Nada, Kobe 657, Japan}
\centerline{email address : UTOSA@JPNYITP.BITNET}
\vskip 2.5cm
\centerline{\large {\bf ABSTRACT} }

\vskip 0.5cm

New geometrical features of the \lgo s
are presented,
for models with a typical type of superpotential.
We show the one-to-one correspondence
between some of the $(a,c)$ states with $U(1)$ charges
$(-1,1)$ and the integral points on the dual polyhedra,
which are useful tools
for the construction of mirror manifolds.
Relying on toric geometry,
these states are shown to correspond to the $(1,1)$ forms
coming from blowing-up processes.
In terms of the above identification,
it can be checked that
the monomial-divisor mirror map for \lg\ orbifolds,
 \ proposed by the author, \
is equivalent to
that mirror map
for \cym s obtained by the mathematicians.
\thispagestyle{empty}

\clearpage

\pagenumbering{arabic}

Mirror symmetry was first discovered in the context of
string compactification \cite{gp,lvw,dixon}.
Due to the relative sign of the two $ U(1) $ charges,
one $ (2,2) $ superconformal field theory allows
two geometrical interpretations,
i.e. topologically distinct \cym s
${\cal M}$ and ${\cal W}$.
Assuming that mirror symmetry is true, some Yukawa couplings
can be determined exactly \cite{cop,hkty1,agm1}.
However, recent analysis of mirror symmetry
is purely geometrical.

Batyrev \cite{vb} proposed a powerful method for constructing
the mirror manifolds of a certain class of \cym s.
He showed that a pair of
$ (\Delta , \Delta^*) $ gives a \cym ,
where $ \Delta $ is a (Newton) polyhedron corresponding to
monomials and $ \Delta^* $ is a dual (or polar) polyhedron
describing the resolution of singularities,
i.e. a point on a one- or two-dimensional face
of $\Delta^*$ corresponds to a $(1,1)$ form
coming from resolution.
Batyrev observed that the exchange of the roles of
$ (\Delta , \Delta^*) $ produces a mirror manifold.

\lg\ models of $ N=2 $ superconformal field theories
are closely related to \cym s
because of their (anti-)chiral ring structures.
If we consider the theory with $ c=9 $,
the $(p,q)$ forms on a \cym\
can be identified with
$(3-p,q)$ states of the $(c,c)$ ring
or $(-p,q)$ states of the $(a,c)$ ring,
where $c \ (a)$ stands for (anti-)chiral
and the states are labeled by the $U(1)$ charges.
These $(c,c)$ and $(a,c)$ rings can be described
in terms of the \lg\ models.

In this paper, we will find a very simple relation
between a $(-1,1)$ state and a point on a face of
$ \Delta^* $,
when a typical type of \lg\ models are considered.
Hence we can identify a $(-1,1)$ state and
a $(1,1)$ form coming from blowing-up processes
in a simple and exact way.
Furthermore, we will show that
the monomial-divisor mirror map for \lgo s proposed in ref.\cite{sa2}
is equivalent to that mirror map of \cym s \cite{agm3,hkty1}.
These are useful extensions of the results
in the previous paper \cite{sa2}.
Our method gives us the possibility
to study the new geometric content
of a class of $N=2$ superconformal field theories.

In this paper, we will restrict our attention to
the superpotential of a form
$W{(X_{i})} =
X_{1}^{a_{1}}+X_{2}^{a_{2}}+X_{3}^{a_{3}}
+X_{4}^{a_{4}}+X_{5}^{a_{5}},$
which corresponds to the Fermat type hypersurface
in $WCP^{4}$.
The \lg\ orbifolds are obtained by
quotienting with an Abelian symmetry group $G$ of
$W{(X_{i})}$,
whose element $g$ acts as an $N \times N$ diagonal matrix,
$g: X_{i} \rightarrow e^{2 \pi i {\tg}}X_{i}$,
where $0 \leq \tg < 1$.
Of course the $U(1)$ twist \
$j: X_{i} \rightarrow e^{2 \pi i {q_{i}}}X_{i}$ \
generates the symmetry group of
$W{(X_{i})}$,
where $q_{i} = {w_{i} \over d}$,\ \
 $W(\la^{w_{i}} X_{i}) = \la^{d}W(X_{i})$ \ and
$\la \in \IC^{\ast}$.
In this paper, we further require that
$ w_{5} = 1 $
since the toric description of
the corresponding \cy\ mirror manifolds
are well-known \cite{vb,hkty1}.

Using the results of Intriligator and Vafa \cite{iv},
we can construct the $(c,c)$ and $(a,c)$ rings.
Also we could have the left and right $U(1)$ charges of
the ground state \
$\keta h$ \
in the $h$-twisted sector
of the $(a,c)$ ring.
In terms of spectral flow, \ $\keta h$ is mapped to the (c,c)
state $\ketc {h^{\prime}}$ with $h^{\prime} = hj^{-1}$.
Then the charges of the (a,c) ground state of $h$-twisted sector
$ \keta {h} $
are obtained to be
\begin{equation}
\label{uac}
\begin{array}{cc}
\left(\begin{array}{c}
J_{0} \\
\bar{J_{0}}
\end{array} \right) &
\end{array}
\keta {h}
=
\begin{array}{cc}
\left(\begin{array}{c}
{ - \sum_{\thd>0}{(1-q_{i}-\thd)}}
+ \sum_{\thd=0} {(2q_{i}-1)}
 \\
{ \sum_{\thd>0}{(1-q_{i}-\thd)}}
\end{array} \right) &
\keta {h}.
\end{array}
\end{equation}

Using this result, we see that the $(-1,1)$ states written in the form
$\keta {j^{l}}$ can always arise from the twisted sector
with $I^{\prime} = 0$
, where $I^{\prime}$ is the number of the invariant fields $X_{i}$
under the $h^{\prime}$ action.
Using the results of ref.\cite{sa1},
it was shown \cite{sa2} that
as long as we consider the \lg\ models
with no or one trivial field,
the $(-1,1)$ states which can be represented by
$\keta {j^{l}}$ may exist only in the twisted sector
with $I^{\prime} = 0$.

Let us turn our attention to geometry.
\cym s are represented by hypersurfaces in $WCP$.
In general, due to the $WCP$ identification\
$z_{i} \sim \la^{w_{i}}z_{i}$,\ \ $\la \in \IC^{\ast}$, \
we have some fixed sets on a hypersurface.
When we consider \cy\ $3$-folds,  possible fixed sets are
fixed points and fixed curves.
To obtain a smooth \cym\, we have to blow up these singularities.

Those \cy\ resolutions can be described in terms of toric geometry
\cite{vb,hkty1}.
Toric geometry describes the structure of a certain class of
geometrical spaces in terms of simple combinatorial data.
To investigate the mirror symmetry,
Batyrev's construction is useful.
We will briefly summarize this method.
Details are presented in \cite{vb,hkty1}.

A (Newton) polyhedron
$  \Delta(w) $ is associated to monomials,
where $w$ means the set of weights $w_i$.
A dual polyhedron
$  \Delta^*(w) $ allows us to describe
the resolution of singularities.
Integral points on faces of dimension one or two of
$  \Delta^*(w) $
correspond to exceptional divisors.
More precisely, points lying on a one-dimensional edge
correspond to exceptional divisors over singular curves,
whereas the points lying in the interior of two-dimensional faces
correspond to the exceptional divisors over singular points.
So, integral points on faces of
$ \Delta^*(w) $ correspond to the
$(1,1)$ forms coming from blowing-up processes.

Since we consider the Fermat type quasihomogeneous polynomial,
the corresponding \cy\ hypersurface consists of
monomials $z_i^{d/w_i}\,\,(i=1,\cdots,5)$.
The associated $4$-dimensional integral convex polyhedron
$\Delta(w)$
is the convex hull of the integral vectors $m$ of the exponents
of all quasi-homogeneous monomials
of degree $d$
shifted by $(-1,\ldots,-1)$, i.e.
$\prod_{i=1}^{5}{z_{i}^{m_{i}+1}}$:
\begin{equation}
\Delta(w):=
\{(m_1,\ldots,m_{5}) \in
\IR^{5}|\sum_{i=1}^{5} w_i m_i=0,m_i\geq-1\} .
\end{equation}
This implies that only the origin is
the point in the interior of $\Delta$.
Its dual polyhedron is defined by
\begin{equation}
\Delta^*=\{\;(x_1,\dots,x_4) \;\vert\;
\sum_{i=1}^4 x_i y_i \geq -1 \;{\rm for \; all \; }
(y_1,\dots,y_4)\in\Delta\;\}.
\end{equation}
In our case it is known that
$ (\Delta , \Delta^*) $
is a reflexive pair.
An $l$-dimensional face
$\Theta \subset \Delta$ can be represented by specifying
its vertices
${\rm v}_{i_1},\cdots,{\rm v}_{i_{k}}$.
Then the dual face $\Theta^*$
is a $(4-l-1)$-dimensional face of $\Delta^*$
and defined by
\begin{equation}
\label{dualface}
\Theta^*=\{ x\in \Delta^*\;
\vert \; (x,{\rm v}_{i_1})=\cdots=(x,{\rm
v}_{i_k})=-1 \},
\end{equation}
where $(*,*)$ is the ordinary inner product.

For our type of models,
we then always obtain as
vertices of $\Delta(w)$
\begin{eqnarray}
&\nu_1& \! \! \! =(d/w_1-1, -1, -1, -1),  \
\nu_2=(-1, d/w_2-1, -1, -1), \
\nu_3=(-1, -1, d/w_3-1, -1),  \nonumber \\
&\nu_4& \! \! \! =(-1, -1, -1, d/w_4-1),  \
\nu_5=(-1,-1,-1,-1),
\end{eqnarray}
and for the vertices of the dual polyhedron
$\Delta^*(w)$ one finds
\begin{eqnarray}
&\nu_1^*&=(1,0,0,0),  \quad
\nu_2^*=(0,1,0,0), \quad
\nu_3^*=(0,0,1,0), \quad
\nu_4^*=(0,0,0,1), \nonumber \\
&\nu_5^*&=(-w_1,-w_2,-w_3,-w_4) .
\end{eqnarray}

For the Fermat type hypersurfaces of degree $d$,
the explicit form of
the monomial-divisor mirror map
has been already studied.
Through this map,
integral points $\mu$ in $\Delta^*(w)$ are mapped to
monomials of the homogeneous coordinates of $WCP^4$ by
\begin{equation}
\label{mdmm}
\mu=(\mu_1,\mu_2,\mu_3,\mu_4)\mapsto
{\prod_{i=1}^4 z_i^{\mu_i d/w_i}
\over\left(\prod_{i=1}^5 z_i\right)^{(\sum_{i=1}^4\mu_{i}) -1}}.
\end{equation}

\vskip 0.5cm

In the following
we will associate an integral point inside $\Delta^*(w)$,
 i.e. an exceptional divisor, with a $(-1,1)$ state
which can be written in the form $\keta {j^{-l}}$.

To explain our method,
we define the phase symmetries $\rho_{i}$ which act on $X_{i}$ as
\begin{equation}
\rho_{i}X_{i} = e^{2 \pi i q_{i} }X_{i},
\end{equation}
with trivial action for other fields.
The operator $ \rho_{i} $
can be represented by a diagonal matrix
whose diagonal matrix elements are 1 except for
$ {(\rho_{i})}_{i,i}
= e^{2 \pi i q_{i} }$.
It is obvious that
\begin{equation}
j = \rho_{1} \cdots \rho_{5}.
\end{equation}

In ref.\cite{sa2} the mirror map
for the $(a,c)$ ground states
in the $j^{-l}$-twisted sector
$\keta {j^{-l}}$
are considered.
In the $j^{-l}$-twisted sector, if a field $X_{i}$ is invariant
then
\begin{equation}
\rho_{i}^{-l} = \rho_{i}^{-l_{i}} = {\rm identity},
\end{equation}
where $-l_{i} \equiv -l \ {\rm mod} \ a_{i}$ and one gets
\begin{equation}
\label{key1}
j^{-l} = \prod_{{-l{q_{i}}} \notin \ZZ} {\rho_{i}^{-l_{i}}}.
\end{equation}
So, we may represent
${\keta {j^{-l}}}
=\keta {\prod_{{-lq_{i}} \notin \ZZ}{\rho_{i}^{-{l_{i}}}}}$.
Furthermore,
we can calculate the $U(1)$ charges of this state
using eq.(\ref{uac})
and the result is
\begin{equation}
{( - \sum_{{-lq_{i}} \notin \ZZ}{l_{i}q_{i}},
\sum_{{-lq_{i}} \notin \ZZ}{l_{i}q_{i}})}.
\end{equation}

The eq.(\ref{key1}) is the key equation for our purpose.
This implies
\begin{equation}
\label{key2}
-lq_i = u_i - l_{i}q_{i} \quad \quad {\rm for} \
i=1 \sim 5,
\end{equation}
where $-l_i$ are defined to be
$-a_i+1 \le -l_{i} \le 0$.
Thus $u_i$ are uniquely determined.
Clearly, $u_i$ are integers
and $u_{i}=0$ if $X_{i}$ is invariant
under $j^{-l}$ action
(see eq.(\ref{key1})).
$u_5$ always vanish
since $w_5 = 1$.

Let $u \equiv (u_1,u_2,u_3,u_4)$
be an integral vector.
In the following we will show
that $u$ is on a face of $\Delta^*(w)$.
So we should assert
that $u$ is just an integral point inside
$\Delta^*(w)$,
which can be identified with the exceptional divisors.
Through this identification,
we obtain the one-to-one correspondence
between the $(-1,1)$ state $\keta {j^{-l}}$
and the exceptional divisor.
Moreover, once this identification is made,
we can see that the monomial-divisor mirror map
for \cym s, i.e. eq.(\ref{mdmm}),
is equivalent to that mirror map
for \lgo s conjectured in ref.\cite{sa2}.

First, we show
that $u$ is a point on a dual face
$\Theta^*$ of $\Delta^*(w)$.
A dual face $\Theta^*$ is specified by some of the vertices
$ \{ \nu_1,\nu_2,\nu_3,\nu_4,\nu_5 \} \subset \Theta$
through the eq.(\ref{dualface}).
So, we have to show
that there are some vectors $\nu_j$ satisfying
\begin{equation}
\label{dual1}
(u,\nu_j) = -1 \quad \quad {\rm {for \ some}} \ j.
\end{equation}
To prove this,
we consider the invariant field $X_j$
under $j^{-l}$ action.
This implies $-lq_{j} = u_{j}$.
It can be shown
that the corresponding vector
$\nu_j$ satisfies eq.(\ref{dual1}).
Denote by ${(\nu_{j})}_{i}$
the $i$-th component of the vector $\nu_j$.
Multiplying $u_i$ by ${(\nu_j)}_{i}$
and using eq.(\ref{key2}), one finds
\begin{equation}
\label{dual2}
(u,\nu_{j}) = - { \sum_{i=1}^{5}{l_{i}q_{i}} },
\end{equation}
where we have used $\sum_{i=1}^{5}{q_{i}} = 1 $, \
$u_{5} = 0$ \ and $-lq_{j} = u_{j}$.

It should be noticed
that the right hand side of eq.(\ref{dual2})
is nothing but the left $U(1)$ charge
of the state $\keta {j^{-l}}$.
Since we consider the $(-1,1)$ state,
it has been proved that eq.(\ref{dual1}) holds
for the vector $\nu_{j}$
with $-lq_{j} = u_{j}$.
Note that $\Theta^*$ is specified by the vectors
${\nu_j} \subset \Theta$ through eq.(\ref{dualface}),
whose corresponding fields $X_{j}$ with the same index
are invariant under $j^{-l}$ action.

Now we turn our attention to
the monomial-divisor mirror map.
As mentioned above,
the monomial-divisor mirror map for \cym s
is summarized in eq.(\ref{mdmm}).
In ref.\cite{sa2}, it was conjectured
that the monomial-divisor mirror map for \lgo s
is obtained to be
\begin{equation}
\label{mdmmlg}
{\keta {\prod_{{-lq_{i}} \notin \ZZ}{\rho_{i}^{-{l_{i}}}}}}
\buildrel {\rm mirror \  pair} \over \longleftrightarrow
{{{\prod_{{-lq_{i}} \notin \ZZ}{\overline{{X_{i}}}^{{l_{i}}}}}}\ket0},
\end{equation}
where $\overline{X_{i}}$ are the fields
of the so-called transposed potential $\overline{W}$
\cite{bh}.
Evidently,
$X_{i} = \overline{X_{i}}$
for our Fermat type potentials.

By identifying the vector $\mu$ in eq.(\ref{mdmm})
with our vector $u$,
we easily obtain
\begin{eqnarray}
\label{mdmmu}
u=(u_1,u_2,u_3,u_4) &\mapsto&
{\prod_{i=1}^4 z_i^{u_i d/w_i}
\over\left(\prod_{i=1}^5 z_i\right)^{ (\sum_{i=1}^4 u_{i}) -1}}
\nonumber
\vspace*{1cm}
\\
&=&
{ {{\prod_{{-lq_{i}} \notin \ZZ}{{z_{i}}^{{l_{i}}}}}} }.
\end{eqnarray}
Since the vector $u$ corresponds to
the $(-1,1)$ state $\keta {j^{-l}}$
and $z_{i}$ to $\overline{X_{i}}$,
one finds that the two monomial-divisor mirror map,
i.e. eq.(\ref{mdmm}) and (\ref{mdmmlg}),
are equivalent.

Note that through the eqs.(\ref{key2}) and (\ref{mdmmu}),
the origin $\nu_{0}^* = (0,0,0,0)$ in $\Delta^*(w)$
corresponds to the $(-1,1)$ state $\keta {j^{-1}}$
and its mirror partner corresponds to
the monomial $z_1z_2z_3z_4z_5$.
They always exist for our type of models.

Let us demonstrate our method
by taking some examples.
Since we have shown $u \in \Theta^*$,
we change our notation of $u$
into $\nu_6^*$
($\nu_7^*, \ \cdots$ \ if several $u$ exist).

First we consider the \lg\ model with the superpotential
\begin{equation}
\label{w21}
W_{1} =
X_{1}^{4}+X_{2}^{4}+X_{3}^{4}+X_{4}^{8}+X_{5}^{8},
\end{equation}
with $U(1)$ charges of $X_{i}$ being
\begin{equation}
( \
\frac{1}{4}, \
\frac{1}{4}, \
\frac{1}{4}, \
\frac{1}{8}, \
\frac{1}{8} \
),
\end{equation}
which were studied in ref.\cite{hkty1,sa2}.
The orbifold model $W_{1} / j$ has a corresponding $\ZZ_2$ fixed curve
which can be written
\begin{equation}
z_{1}^{4}+z_{2}^{4}+z_{3}^{4} = 0 \ \ \
{\rm in } \  WCP_{(1,1,2,2,2)}^{4}.
\end{equation}
In this model, we have one twisted ground state
$\keta {j^{-4}}$
which corresponds to the resulting one $(1,1)$ form
after blowing-up.
Since
${j^{-4}} = {\rho_{4}^{-4}}{\rho_{5}^{-4}}$,
we can calculate
$\nu_{6}^* = (-1,-1,-1,0).$
Also it can be seen
that $\nu_{6}^*$ is on the dual face $\Theta^*$
specified by the vectors
$\nu_{1},\nu_{2},\nu_{3}$
(in the following we denote by $\Theta^*(1,2,3)$
this face).
{}From eq.(\ref{mdmmlg}),
$\nu_6^*$ is mapped to the monomial
$\overline{X_{4}^{4}}\overline{X_{5}^{4}}$.
These are the same results
as the ones obtained in \cite{hkty1}.

As a more complicated example, we take the following
\lg\ superpotential
\begin{equation}
\label{w63}
W_{2} =
X_{1}^{3}+X_{2}^{3}+X_{3}^{6}+X_{4}^{9}+X_{5}^{18},
\end{equation}
with $U(1)$ charges
\begin{equation}
( \
\frac{1}{3}, \
\frac{1}{3}, \
\frac{1}{6}, \
\frac{1}{9}, \
\frac{1}{18} \
).
\end{equation}
This model is considered in \cite{agm1,sa2}.
The orbifold model $W_{2} / j$ has one corresponding
$\ZZ_2$ fixed curve,
one corresponding $\ZZ_3$ fixed curve
and  corresponding $\ZZ_6$ fixed points on the
intersections of these curves.
They can be written as
\begin{equation}
{\rm \ZZ_{2} \ fixed \ curve} \
z_{1}^{3}+z_{2}^{3}+z_{4}^{9} = 0
\end{equation}
\begin{equation}
{\rm \ZZ_{3} \ fixed \ curve} \
z_{1}^{3}+z_{2}^{3}+z_{3}^{6} = 0
\end{equation}
\begin{equation}
{\rm \ZZ_{6} \ fixed \ points} \
z_{1}^{3}+z_{2}^{3} = 0 \ \quad
{\rm in } \ WCP_{(6,6,3,2,1)}^{4}.
\end{equation}

After blowing-up,
one obtains four $(1,1)$ forms
whose mirror partners
can be associated to monomial deformation.
It is easy to find the corresponding
$(-1,1)$ states
written in the form $\keta {j^{-l}}$,
the vectors $\nu_{j}^*$,
dual faces $\Theta^*$ on which $\nu_j^*$ are lying,
and mirror partners.
The results are displayed in Table \ref{mdl2},
 where we have omitted the bar over $X_{i}$.

\begin{table}[htbp]
\[ \begin{tabular}{||c|c|c|c||} \hline
(a,c) state & $\nu_j^*$ vector & dual face & mirror partner \\ \hline \hline
$\keta {j^{-3}}$ &
$\nu_6^*=(-1,-1,0,0)$ &
$\Theta^*(1,2)$ &
${{X_{3}}^{3}}
{{X_{4}}^{3}}
{{X_{5}}^{3}}
\ket0 $ \\ \hline
$\keta {j^{-6}}$ &
$\nu_7^*=(-2,-2,-1,0)$ &
$\Theta^*(1,2,3)$ &
${{X_{4}}^{6}}
{{X_{5}}^{6}}
\ket0 $ \\ \hline
$\keta {j^{-9}}$ &
$\nu_8^*=(-3,-3,-1,-1)$ &
$\Theta^*(1,2,4)$ &
${{X_{3}}^{3}}
{{X_{5}}^{9}}
\ket0 $ \\ \hline
$\keta {j^{-12}}$ &
$\nu_9^*=(-4,-4,-2,-1)$ &
$\Theta^*(1,2,3)$ &
${{X_{4}}^{3}}
{{X_{5}}^{12}}
\ket0 $ \\ \hline
\end{tabular} \]
\caption{The monomial-divisor mirror map for
\lg\ orbifolds of $W_{2}$}
\label{mdl2}
\end{table}
This result agrees with the one obtained in \cite{agm1} .
Note that we do not need any geometrical informations
such as the number of fixed sets or the relations among them.

In this model, there are two $(-1,1)$ states represented by
$X_{1} {\keta {j^{-2}}}$ \ \
and $X_{2} {\keta {j^{-2}}}$.
The mirror partners for the corresponding $(1,1)$ forms
cannot be described by monomials.
Unfortunately, we have not succeeded in associating
these states with toric data yet.

\vskip0.5cm

In conclusion, we have discovered
a simple and direct connection
between the \lg\ and  the toric
descriptions of our class of \cym s.
In other words,
we find that a \lgo\ and a corresponding \cym\
have just the same toric data,
as far as models with a typical type of superpotentials
are concerned.
Our method enables us to calculate toric data easily
without referring to
the geometrical nature of $\Delta^*(w)$.
We can uniquely identify
a $(-1,1)$ state $\keta {j^{-l}}$
with a $(1,1)$ form resulting from resolution.
This would be a useful technique
for analyzing the Yukawa couplings,
especially when \cym s with several $(1,1)$ forms
are considered.

The Fermat type superpotential considered in this paper
corresponds to the Gepner model of A-type \cite{g1}.
However, there are non-Fermat-type potentials with
only the singularities which can be treated through toric geometry.
For example, there is the hypersurface embedded in
$WCP_{(1,2,2,2,3)}^{4}$ ,
which gets two toric divisors after blowing up \cite{beka}.
For this model, our method
for the calculation of the vector $u$
could not be applied as it is.
Some extensions are needed.
We will report it elsewhere \cite{sa4}.

\vspace{1cm}

{\it Acknowledgements} :
The author would like to thank C.S. Lim
for careful reading of this manuscript.



\newpage

\begin{thebibliography}{99}
\bibitem{gp} B.R. Greene and M.R. Plesser, Nucl. Phys.
{\bf B338} (1990) 15.
\bibitem{lvw} W. Lerche, C. Vafa and N.P. Warner, Nucl. Phys.
{\bf B324} (1989) 427.
\bibitem{dixon} L. Dixon, in
{\it Superstrings, Unified Theories and Cosmology 1987},
G. Furlan et.al.,eds., World Scientific, Singapore (1988).
\bibitem{cop} P. Candelas, X. de la Ossa, P.S. Green
and L. Parkes,
Phys. Lett. {\bf B258} (1991) 118 and Nucl. Phys.
{\bf B359} (1991) 21.
\bibitem{hkty1} H. Hosono, A. Klemm, S. Theisen and S.-T. Yau,
Commun. Math. Phys. {\bf 167} (1995) 301.
\bibitem{agm1} P.S. Aspinwall, B.R. Greene and D. Morrison,
Phys. Lett. {\bf B303} (1993) 249
, Nucl. Phys. {\bf B416} (1993) 414.
\bibitem{vb} V. Batyrev, J. Alg. Geom. {\bf 3} (1994) 493.
\bibitem{sa2} H. Sato, Mod. Phys. Lett. {\bf A9} (1994) 3721.
\bibitem{agm3} P.S. Aspinwall, B.R. Greene
and D. Morrison, Int. Math. Res. Notices (1993) 319.
\bibitem{iv} K. Intriligator and C. Vafa, Nucl. Phys.
{\bf B339} (1990) 95;
C. Vafa, Mod. Phys. Lett. {\bf A4} (1989) 1169.
\bibitem{sa1} H. Sato, Mod. Phys. Lett. {\bf A9} (1994) 885.
\bibitem{bh} P. Berglund and T. H\"ubsch,
Nucl. Phys. {\bf B 393} (1993) 377.
\bibitem{g1} D. Gepner, Phys. Lett. {\bf B199} (1987) 380
, Nucl. Phys. {\bf B296} (1988) 757.
\bibitem{beka} P. Berglund and S. Katz,
to appear in {\it Essays on Mirror Manifolds II}
(Ed. S.-T. Yau), hep-th/9406008
\bibitem{sa4} H. Sato, in preparation.
\end{thebibliography}
\end{document}